\documentclass[final,3p,sort&compress]{elsarticle}

\usepackage{graphicx}
\usepackage{amsmath,amssymb,amsfonts}
\usepackage{bm}
\def\be{\begin{equation}}
\def\ee{\end{equation}}
\def\ba{\begin{eqnarray}}
\def\ea{\end{eqnarray}}
\def\bd{\begin{displaymath}}
\def\ed{\end{displaymath}}
\def\bq{\begin{eqnarray}}
\def\eq{\end{eqnarray}}

\begin{document}

\begin{frontmatter}



\title{Full quantum mechanical analysis\\ of atomic three-grating
Mach-Zehnder interferometry}


\author[label1]{A. S. Sanz\corref{corresp}} \ead{asanz@iff.csic.es}
\author[label2]{M. Davidovi\'c}
\author[label3]{M. Bo\v zi\'c}

\cortext[corresp]{Corresponding author}

\address[label1]{Instituto de F\'{\i}sica Fundamental (IFF-CSIC),
Serrano 123, 28006 Madrid, Spain}

\address[label2]{Faculty of Civil Engineering, University of Belgrade,
Bulevar Kralja Aleksandra 73, 11000 Belgrade, Serbia}

\address[label3]{Institute of Physics, University of Belgrade,
Pregrevica 118, 11080 Belgrade, Serbia}

\begin{abstract}
Atomic three-grating Mach-Zehnder interferometry constitutes an
important tool to probe fundamental aspects of the quantum theory.
There is, however, a remarkable gap in the literature between the
oversimplified models and robust numerical simulations considered to
describe the corresponding experiments.
Consequently, the former usually lead to paradoxical scenarios, such
as the wave-particle dual behavior of atoms, while the latter make
difficult the data analysis in simple terms.
Here these issues are tackled by means of a simple grating working
model consisting of evenly-spaced Gaussian slits.
As is shown, this model suffices to explore and explain such experiments
both analytically and numerically, giving a good account of the full
atomic journey inside the interferometer, and hence contributing to make
less mystic the physics involved.
More specifically, it provides a clear and unambiguous picture of the
wavefront splitting that takes place inside the interferometer,
illustrating how the momentum along each emerging diffraction order
is well defined even though the wave function itself still displays
a rather complex shape.
To this end, the local transverse momentum is also introduced in this
context as a reliable analytical tool.
The splitting, apart from being a key issue to understand atomic
Mach-Zehnder interferometry, also demonstrates at a fundamental level
how wave and particle aspects are always present in the experiment,
without incurring in any contradiction or interpretive paradox.
On the other hand, at a practical level, the generality and versatility
of the model and methodology presented, makes them suitable to attack
analogous problems in a simple manner after a convenient tuning.
\end{abstract}

%%%%%%%%%%%%%%%%%%%%%%%%%%%%%%%%%%%%%%%%%%%%%%%%%%%%%%%%%%%%%%%%%%%%%%%


\end{frontmatter}


%%%%%%%%%%%%%%%%%%%%%%%%%%%%%%%%%%%%%%%%%%%%%%%%%%%%%%%%%%%%%%%%%%%%%%%
%%%%%%%%%%%%%%%%%%%%%%%%%%%%%%%%%%%%%%%%%%%%%%%%%%%%%%%%%%%%%%%%%%%%%%%

\section{Introduction}
\label{sec1}

Matter-wave interferometry constitutes an important application of
quantum interference with both fundamental and practical interests
\cite{badurek,adams,berman,cronin,arndt1,arndt5}.
In analogy to optics, this sensitive technique allows us to determine
properties of the diffracted particles as well as of any other element
acting on them during their transit through the interferometer.
It was at the beginning of the 1990s when Kasevich and Chu
\cite{kasevich:PRL:1991-2} showed that matter-wave Mach-Zehnder
interferometry can be achieved by using the same basic ideas of its
optical analog: if the atomic wave function can be coherently split up,
and later on each diffracted wave is conveniently redirected in order
to eventually achieve their recombination on some space region, then an
interference pattern will arise on that spot.
For neutral atoms this can be done by means of periodic gratings,
which play the role of optical beam splitters.
This property, exploited in different diffraction experiments with
fundamental purposes \cite{stern,gould,kapitza,batelaan,keith1}, gave
rise to the former experimental implementations of atomic Mach-Zehnder
interferometers in the early 1990s, first with transmission gratings
\cite{keith2} and then with light standing waves \cite{rasel}.
These interferometers are based on a very efficient production of
spatially separated coherent waves \cite{carnal2}.
Similar interferometers are also used for large molecules
\cite{zeilinger3,arndt3,arndt4}, although due to their relatively
smaller thermal wavelength, they work within the near field or Fresnel
regime, benefiting from the grating self-imaging produced by the
so-called Talbot-Lau effect \cite{clauser1,clauser2}, a combination
of the Talbot \cite{talbot,rayleigh,winthrop,latimer} and
Lau \cite{lau,lohmann,sudol} effects.

\begin{figure}[t]
 \begin{center}
 \includegraphics[width=8cm]{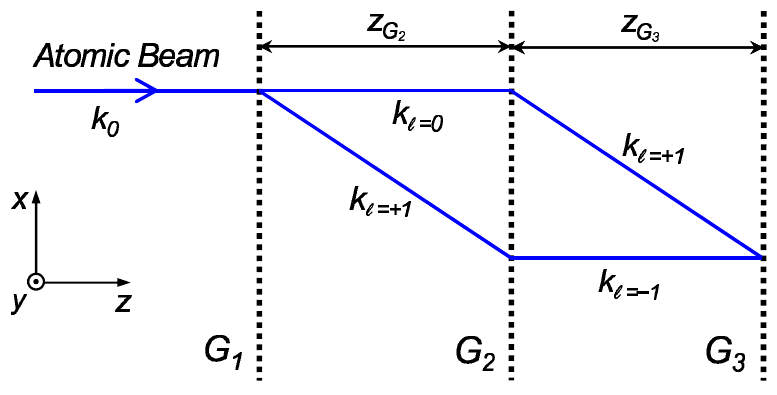}
 \caption{\label{fig1}
  Sketch of an atomic three-grating Mach-Zehnder interferometer.
  This class of interferometers consists of three equally-spaced
  ($z_{G_2} = z_{G_3}$), parallel gratings, namely $G_1$, $G_2$, and $G_3$.
  These gratings play the role of the set of beam splitters ($G_1$ and
  $G_3$) and mirrors ($G_2$) of a conventional optical interferometer.}
 \end{center}
\end{figure}

To describe and analyze this kind of experiments different
analytical and numerical treatments have been proposed in the
literature \cite{adams,berman,cronin,storey:JPIIF:1994}.
Nonetheless, there is a substantial gap between simple models and exact
numerical simulations, which makes difficult getting a unified view of
the physics involved in these experiments.
Consequently, many times our understanding of them is oversimplified,
which leads us to emphasize ``paradoxical'' aspects of quantum
mechanics.
One of these aspects is, for example, the commonly assumed
wave-particle dual nature of quantum systems.
Depending on how the experiment is performed, one ``decides'' between
one or the other, which manifests as one or another type of outcome.
Now, leaving aside ontological issues, a pragmatic, accurate description
of the experiment requires the use of a wave function.
The evolution of this wave function in the course of the experiment is
strongly influenced by the boundary conditions associated with such an
experiment, as well as by any other physical effect that might take
place (e.g., presence of photon scattering events \cite{cronin}), which
will unavoidably lead to different outcomes.
That is, any trace of paradox disappears if we just focus on the wave
function and all the factors that affect it during the performance of
the experiment --- we obtain what should be obtained, leaving not much
room for speculating about dual behaviors.
This is precisely the scenario posed by atomic Mach-Zehnder
interferometers, where it is common to describe the evolution between
consecutive gratings in terms of classical-like paths (see
Fig.~\ref{fig1}), although their recombination (and, actually, also
their emergence) has to do with a pure wave-like behavior.

Moved by these facts and their important implications, here we revisit
the problem with a working model consisting of a set of three gratings
with Gaussian slits (slits characterized by a Gaussian transmission
function), while its analysis is conducted by means of a combination
of the position and momentum representations of the (atomic) wave
function.
As is shown, the synergy between analytical results and numerical
simulations obtained in this way helps to describe and understand the
functioning of these interferometers in a relatively simple manner.
Specifically, we provide a clear picture of the wavefront splitting
process that takes place at the gratings, showing how the typical
path-like picture of the interferometer \cite{nachman} (see
Fig.~\ref{fig1}) coexists with the complex interference patterns
exhibited by the wave function between consecutive gratings.
This is a key point to understand the simplified models commonly used
in the literature to explain this type of interferometers, where the
particular shape of the wave function is neglected and only the momentum
carried along the paths associated with each involved diffraction order is
considered.
In this regard, we have introduced the concept of local transverse
momentum, borrowed from the Bohmian formulation of quantum mechanics
\cite{asanz-bk1}, as analytical tool.
By means of this quantity it is possible to properly quantify the local
value of the momentum (not to be confused with the usual momentum
expectation value) at each point of the transverse coordinate, which
is related to the quantum flux \cite{schiff-bk,zhang-bk} evaluated on
that point.
Furthermore, we would also like to stress the practical side of this
model as an efficient tool to attack analogous problems with presence
of incoherence sources and/or decoherent events in a simple manner.

This work has been organized as follows.
The general theoretical elements involved in our analysis of the
atomic three-grating Mach-Zehnder interferometer are described
in Sec.~\ref{sec2}, including the Gaussian grating model considered.
Analytical results obtained from this model in the far field are
presented and discussed in Sec.~\ref{sec4}.
The outcomes from the numerical simulations illustrating different
aspects of the wave-function full evolution between consecutive
gratings are analyzed and discussed in Sec.~\ref{sec5}.
The conclusions from this work are summarized in Sec.~\ref{sec6}.

%%%%%%%%%%%%%%%%%%%%%%%%%%%%%%%%%%%%%%%%%%%%%%%%%%%%%%%%%%%%%%%%%%%%%%%
%%%%%%%%%%%%%%%%%%%%%%%%%%%%%%%%%%%%%%%%%%%%%%%%%%%%%%%%%%%%%%%%%%%%%%%

\section{Theory}
\label{sec2}

%%%%%%%%%%%%%%%%%%%%%%%%%%%%%%%%%%%%%%%%%%%%%%%%%%%%%%%%%%%%%%%%%%%%%%%

\subsection{General aspects of grating diffraction}
\label{sec21}

Atomic three-grating Mach-Zehnder interferometers consist of three
evenly-spaced and parallel periodic gratings, as it is illustrated in
Fig.~\ref{fig1}.
In this sketch the slits are parallel to the $y$ axis and span a
relatively long distance (larger than the cross-section of the
incident beam).
This implies translational invariance along the $y$ axis, which makes
diffraction to be independent of this coordinate and allows a reduction
of the problem dimensionality to the transverse ($x$) and longitudinal
($z$) directions.
Under realistic experimental working conditions, taken from the former
1991 experiment by Keith et al.\ \cite{keith2} (see Sec.~\ref{sec22}),
the propagation of the atomic beam along the longitudinal direction is
typically faster than its translational spreading \cite{sanz-jpa}.
From the experimental data reported in \cite{keith2}, we notice that
the distance between consecutive gratings is of the order of half a
meter, while the separation between consecutive diffraction orders is
about a few tenths of microns.
These are just paraxial conditions, which allow us to decouple the
transverse and longitudinal (translational) degrees of freedom for
practical purposes.
This assumption not only simplifies the
analytical treatment, but also provides us with a neater dynamical
picture of the emergence of well-resolved diffraction orders and
their subsequent recombination (see Sec.~\ref{sec5}), avoiding the
complexities involved by reflections at the gratings
\cite{sanz-ch-micha}.
Also for simplicity, short-range interactions between the diffracted
particles and the grating \cite{sanz-ch-micha,miller:JCP-2:2001}, as
well as imperfections or thermal effects associated with the latter
\cite{arndt3,grisenti1,grisenti2}, are neglected in this work, although
they can be easily implemented by varying some of the parameters of the
model (see discussion in this regard in Sec.~\ref{sec22}).

The atoms leaving the source follow a Boltzmann velocity distribution
at a temperature $T$, which implies an associated average thermal de
Broglie wavelength $\lambda = 2\pi\hbar/\sqrt{3mk_BT}$.
Although the beam is not fully monochromatic under these conditions,
the presence of two consecutive collimating slits beyond the source
leads to a nearly plane (monochromatic) wavefront behind the second
of these slits.
This is the beam impinging on $G_1$.
In spite of possible deviations from full monochromaticity, for
practical purposes and analytical simplicity we assume that this
beam is nearly monochromatic.
Accordingly, if $k^2 = k_x^2 + k_z^2$, where $k=2\pi/\lambda$, we can
assume that $k_z \gg k_x$ and $k \approx k_z$.
This allows us to express the atomic wave function at any time as a
product state,
\be
 \Psi(x,z,t) \approx \psi(x,t) e^{ik_z z - iE_z t/\hbar} .
 \label{eq0}
\ee
The longitudinal component is a plane wave with average momentum
$p_z = \hbar k_z$ and energy $E_z = p_z^2/2m = \hbar k_z^2/2m$.
The transverse component, $\psi(x,t)$, is a solution of the
time-dependent, free-particle Schr\"odinger equation
\be
 i\hbar\ \frac{\partial \psi}{\partial t}
 = -\frac{\hbar^2}{2m} \frac{\partial^2 \psi}{\partial x^2} ,
 \label{eqschro}
\ee
with initial condition $\psi(x,0)$.

Solutions of Eq.~(\ref{eqschro}) can be readily found by computing
the Fourier transform of $\psi(x,t)$ \cite{asanz-bk1},
\be
 \psi(x,t) = \frac{1}{\sqrt{2\pi}}\!
  \int \tilde{\psi}(k_x,t) e^{ik_x x} dk_x .
 \label{eqmrept}
\ee
Physically this is just a way to recast $\psi(x,t)$ as a linear
combination or superposition of plane waves, $e^{ik_x x}$, each one
contributing with a weight and phase specified by
$\tilde{\psi}(k_x,t)$.
Notice that $\tilde{\psi}(k_x,t)$ is just the representation of the
wave function in the momentum (reciprocal) space, which depends on
the reciprocal variable or momentum $k_x$.
Substituting (\ref{eqmrept}) into Eq.~(\ref{eqschro}) leads to
\be
 i\hbar\ \! \frac{\partial \tilde{\psi}}{\partial t}
 = \left( \frac{\hbar^2 k_x^2}{2m}\right) \tilde{\psi} ,
 \label{eqschrop}
\ee
which, after integration in time, yields
\be
 \tilde{\psi}(k_x,t) = \tilde{\psi}(k_x,0) e^{-i\hbar k_x^2 t/2m} .
 \label{eqwfmomt}
\ee
The initial condition $\tilde{\psi}(k_x,0)$
corresponds to the representation of the initial wave function,
$\psi(x,0)$, in the momentum space,
\be
 \tilde{\psi}(k_x,0) = \frac{1}{\sqrt{2\pi}}\!
   \int \psi(x,0) e^{-ik_x x} dx .
 \label{eqmrep0}
\ee
The solution (\ref{eqwfmomt}), with initial condition (\ref{eqmrep0}),
allows us to rearrange Eq.~(\ref{eqmrept}) as
%
%\ba
% \psi(x,t) & = & \frac{1}{\sqrt{2\pi}}
% \int \tilde{\psi}(k_x,0) e^{ik_x x - i\hbar k_x^2 t/2m} dk_x ,
% \nonumber \\
% & = & \frac{1}{2\pi}
% \iint \psi(x',0) e^{ik_x (x-x') - i\hbar k_x^2 t/2m} dk_x dx' .
% \nonumber \\ & &
% \label{eqmrept2}
%\ea
%
%This expression can be simplified by integrating in $k_x$, which renders
%
\be
 \psi(x,t) = \sqrt{\frac{m}{2\pi i\hbar t}}
  \int \psi(x',0) e^{im(x-x')^2/2\hbar t} dx' .
 \label{eqmrept3}
\ee
after integrating in $k_x$.
Notice in this latter expression that the quantity
\be
 \mathcal{K}(x,x') \equiv \sqrt{\frac{m}{2\pi i\hbar t}}\ \!
   e^{im(x-x')^2/2\hbar t}
 \label{kernel}
\ee
corresponds to the free-particle kernel or propagator, which can be
alternatively obtained by means of a rather longer way using path
integrals \cite{feynman-hibbs}.

Given that the longitudinal component of the wave function is a plane
wave that propagates with velocity $v = \hbar k_z/m \approx \hbar k/m$,
%
%\be
% v = \frac{\hbar k_z}{m} \approx \frac{\hbar k}{m} ,
% \label{vperp}
%\ee
%
the problem can be reparameterized in terms of the longitudinal
coordinate, $z$.
That is, the wave function (\ref{eqmrept3}) describing the wave
function evolution along the transverse direction (accounted for by
the $x$ coordinate) can be recast in terms of this coordinate,
\be
 \psi(x,z) \approx \sqrt{\frac{k}{2\pi iz}}
  \int \psi(x',0) e^{ik(x-x')^2/2z} dx' ,
 \label{e14}
\ee
with $z = vt \approx (\hbar k/m) t$.
%
%\be
% z = vt \approx \left(\frac{\hbar k}{m}\right) t .
% \label{zvst}
%\ee
%
This form is mass-independent and therefore can be used advantageously
to describe massive particles as well as light (when the latter
is described by a scalar field \cite{davidovic:PhysScr:2010}).

The difference between various problems is established by the initial
condition $\psi(x,z_0)$, where $z_0$ denotes the starting position
along the $z$ axis from which the (transverse) wave function starts
its propagation.
If the gratings are assumed to be infinitesimally thin along the
longitudinal direction, we can consider a simple relationship between
the ansatz and the incident wave function:
\be
 \psi_G(x,z_0) = \hat{T}_G \{\psi_{\rm inc}^G(x,z_0)\} ,
 \label{e24}
\ee
where $\psi_G(x,z_0)$ is the ansatz (diffracted) wave function just
behind the grating $G$ (located at $z_0$), $\hat{T}_G$ is the
transmission operator characterizing $G$, and $\psi_{\rm inc}^G(x,z_0)$
is the wave function incident onto $G$.
The transmission operator plays here a role analogous to that of the
scattering operator or S-matrix in scattering theory
\cite{ballentine-bk};
the details of its action on the incident wave function are described
below in the context of Gaussian-slit gratings.
Now, taking into account Eq.~(\ref{e24}), the propagation between two
consecutive gratings, given by Eq.~(\ref{e14}), acquires the functional
form:
\be
 \psi_G(x,z) = \sqrt{\frac{k}{2\pi iz}}\!
  \int \hat{T}_G\{\psi_{\rm inc}^G(x',z_0)\} e^{ik(x-x')^2/2z} dx' .
 \label{e14b}
\ee
For practical purposes, and without loss of generality, later on the
problem will be piecewise solved, particularizing Eq.~(\ref{e14b}) to
the transit from $G_1$ to $G_2$ ($\psi_{G_1}$) and from $G_2$ to $G_3$
($\psi_{G_2}$).
For both transits the origin will be set at the corresponding grating
(i.e., $z_0=0$ in both cases).
As for the incident wave functions, $\psi_{\rm inc}^{G_1}(x,0)$ will be
an incoming plane wave, while
$\psi_{\rm inc}^{G_2}(x,0) = \psi_{G_1}(x,z_{G_2})$.

In general the integral (\ref{e14b}) cannot be solved analytically
except for Gaussian states.
Nonetheless, in the far field ($f\!f$) or Fraunhofer regime it acquires
a simpler form and admits some additional analytical solutions \cite{dimic}.
In this regime, we have $x \gg x'$ and therefore the phase factor of
Eq.~(\ref{e14}) can be approximated by
\be
 \frac{k(x-x')^2}{2z} \approx \frac{kx^2}{2z} - \frac{kxx'}{z} .
 \label{ffc}
\ee
This allows to recast Eq.~(\ref{e14}) as
\be
 \psi_G^{f\!f}(x,z) \approx \sqrt{\frac{k}{2\pi iz}}\ \! e^{ikx^2/2z}\!
   \int \psi(x',0) e^{-ik_x x'} dx' ,
 \label{e25}
\ee
where we introduce the definition $k_x = kx/z$, with $x/z$ being the
direction cosine \cite{lipson3-bk} with respect to the origin at the
grating $G$ (another physical explanation will be provided for this
choice of identifying $k_x$ with a transverse momentum; see
Sec.~\ref{sec51}).
Comparing this integral with (\ref{eqmrep0}), we find that except
for a phase factor the wave function in the far field is proportional
to the representation of the initial wave function in the momentum
space:
\be
 \psi_G^{f\!f}(x,z) \approx \sqrt{\frac{k}{iz}}\
  \tilde{\psi}_G(k_x) e^{ikx^2/2z} .
 \label{e26}
\ee
Fraunhofer diffraction can then be understood as the Fourier image of the
initial wave function.
Physically, this means that in the far field (asymptotically) the
global shape of the probability density is independent of the distance
from the grating, mimicking the transverse momentum distribution:
$|\psi_G^{f\!f}(x,z)|^2 \propto |\tilde{\psi}_G(k_x)|^2$.
In other words, the aspect ratio of the wave function remains invariant
with $z$ or, equivalently, with time.
Furthermore, because $\tilde{\psi}_G(k_x)$ is related to the grating
transmission, the far-field wave function is just a manifestation of the
grating structure, thus providing us with information about it and not
only about properties of the diffracted atom.
Note the analogy with optics \cite{hecht-bk}, where the wave
amplitude in the far field or Fraunhofer regime is just the Fourier
transform of the aperture function evaluated at a spatial frequency
precisely given \mbox{by $k_x$.}

Regarding the phase factor that appears in Eq.~(\ref{e26}), it has not
been recast in terms of $k_x$ on purpose, because of its quadratic
dependence on $x$.
As it can be noticed, if we apply the usual (transverse) momentum
operator, $\hat{p}_x = -i\hbar\partial/\partial x$ to (\ref{e26}),
we find
%
%\ba
% \hat{p}_x \psi_G^{f\!f} & \approx & \sqrt{\frac{k}{iz}}
%  \left[ \frac{\hbar kx}{z}\ \tilde{\psi}_G
%  - \frac{i\hbar k}{z} \frac{\partial \tilde{\psi}_G}{\partial k_x}
%  \right] e^{ikx^2/2z} \nonumber \\
% & = & \sqrt{\frac{k}{iz}} \frac{\hbar k}{z}
%  \left[ \int (x - x') \psi(x',0) e^{-ik_x x'} dx'
%  \right] e^{ikx^2/2z} \nonumber \\
% & \approx & \frac{\hbar kx}{z}\ \psi_G^{f\!f}
% = \hbar k_x \psi_G^{f\!f} ,
%\ea
%
\ba
 \hat{p}_x \psi_G^{f\!f} & \propto &
  \left( \frac{\hbar kx}{z}\ \tilde{\psi}_G
  - \frac{i\hbar k}{z} \frac{\partial \tilde{\psi}_G}{\partial k_x}
  \right) e^{ikx^2/2z} \nonumber \\
 & = & \frac{\hbar k}{z}
  \left[ \int (x - x') \psi(x',0) e^{-ik_x x'} dx'
  \right] e^{ikx^2/2z} \nonumber \\
 & \propto & \frac{\hbar kx}{z}\ \psi_G^{f\!f}
 = \hbar k_x \psi_G^{f\!f} ,
\ea
where we have considered the equality $\partial/\partial x =
(\partial k_x/\partial x) \partial/\partial k_x =
(k/z)\partial/\partial k_x$ in the first line,
and the approximation $x \gg x'$ in the last step.
Hence, to some extent it can be said that the far-field wave function
evolves locally (at each point) as an effective plane wave with
(effective) momentum $p_x = \hbar kx/z = \hbar k_x$ (see
Sec.~\ref{sec51}).
Taking this into account, Eq.~(\ref{e26}) can be recast in terms of
plane waves, as
%
%\be
% \psi_G^{f\!f}(x,z) \approx \sqrt{\frac{k}{iz}}\
%  \tilde{\psi}_G(k_x) e^{ik_x x} ,
% \label{e26b}
%\ee
%
\be
 \psi_G^{f\!f}(x,z) \propto \tilde{\psi}_G(k_x) e^{ik_x x} .
 \label{e26b}
\ee
%
%without loss of generality.

%%%%%%%%%%%%%%%%%%%%%%%%%%%%%%%%%%%%%%%%%%%%%%%%%%%%%%%%%%%%%%%%%%%%%%%
%%%%%%%%%%%%%%%%%%%%%%%%%%%%%%%%%%%%%%%%%%%%%%%%%%%%%%%%%%%%%%%%%%%%%%%

\subsection{The Gaussian grating model}
\label{sec22}

In order to get a more quantitative idea on the amplitude splitting
process and the subsequent recombination, now we are going to introduce
a simple model consisting of three identical periodic gratings with
Gaussian slits.
By Gaussian slit we just mean a slit characterized by a Gaussian
transmission function \cite{feynman-hibbs,sanz2}, i.e., a transmission
which is maximum at the center of the slit and decreases smoothly in a
Gaussian fashion towards the slit boundaries.
This behavior can be observed, for example, in situations
where the problem is tackled from a scattering viewpoint and the
interaction between the slit and the diffracted particle is modeled
by means of a realistic interaction soft potential
\cite{sanz-ch-micha,miller:JCP-2:2001}.
Gaussian transmissions are also in compliance with the fact that the
incident beam is not fully monochromatic (see Sec.~\ref{sec2}), which
leads to a relatively fast decay (actually, in a Gaussian fashion)
of eventual diffraction orders as we move apart from the incident
propagation direction (in agreement with real matter wave experimental
observations).
As it will be seen, this model is very convenient both analytically
and numerically.
Regarding the value of the different physical parameters, we have
considered without loss of generality those reported by Keith
{\it et al.}~\cite{keith2} in their former experiment on three-grating
Mach-Zehnder interferometry with sodium atoms.
Accordingly, the thermal de Broglie wavelength of the sodium atoms is
$\lambda = 16$~pm ($k = 0.393$~pm$^{-1}$), the grating period is
$d=0.4$~$\mu$m, the slit width is $w = 0.2$~$\mu$m, and the distance
between two consecutive gratings is $L = 0.663$~m.

The action of the transmission operator on an incident wave function
is modeled as
\be
 \hat{T}_G \{\psi_{\rm inc}\} =
  \sum_{j=1}^N \beta_j\ \! e^{-(x - x_j)^2/4\sigma_0^2 + ik_{j,x0}x} ,
 \label{transmit}
\ee
i.e., each slit produces a Gaussian diffracted wave with a width
$\sigma_0=w/2$, such that it covers an effective distance of $d/2$
(it is almost zero at $x = x_j \pm w/2$).
If $\psi_{\rm inc}^G$ is not a plane wave, the amount of probability
transmitted through each slit will be different.
In order to account for this fact, we introduce an {\it opacity}
parameter $\beta_j$ specifying how much the $j$th slit (i.e., the slit
centered at $x_j$) contributes to the total diffracted wave function.
In particular, here we have considered
$\beta_j = |\psi_{\rm inc}^G(x_j,z_0)|/\sqrt{{\rm max}\{|\psi_{\rm inc}^G(x,z_0)|^2\}}$.
In this contexts, $N$ must be understood as the total, effective
number of slits that contribute to the diffracted wave (we have
neglected contributions from slits such that their associated
$\beta$ are below a certain onset).
Finally, it can also happen that the transverse momentum varies
locally along $\psi_{\rm inc}^G$.
This effect is taken into account by associating a momentum $k_{j,x0}$
with each Gaussian wave, with its value being determined by the local
momentum of $\psi_{\rm inc}^G$ at the center of the $j$th slit.
As shown below, this model is very convenient both analytically and
computationally.

In order to show how the model works in a simple case, consider that
the wave function incident onto $G_1$ is a plane wave given by
\be
 \psi_{\rm inc}^{G_1}(x,0) \sim e^{ik_{x0} x} ,
 \label{e38}
\ee
where $k_{x0} = k_0 \sin \theta$, with $\theta$ being the angle between
the $z$ axis and the direction of the incident wave vector (later on we
will particularize to the case of normal incidence, so that $k_{x0}=0$).
For example, according to the experiment, the passage of the sodium
atoms through two 20-$\mu$m--slits produces a collimated beam
of about 20~$\mu$ width, which covers about 50 slits at $G_1$.
In this sense Eq.~(\ref{e38}) constitutes a reliable guess.
After its substitution into (\ref{transmit}), with $\beta_j=1$ and
$k_{j,x0}=k_{x0}$ for all $j$, we obtain the ansatz (\ref{e24}) behind
$G_1$ (i.e., the initial, diffracted wave function), which is a
coherent superposition of Gaussian wave packets,
\be
 \psi_{G_1}(x,0) \approx \frac{1}{\sqrt{N}} \sum_{j=1}^N
  \left(\frac{1}{2\pi\sigma_0^2}\right)^{1/4}
  e^{-(x - x_j)^2/4\sigma_0^2 + ik_{x0} x} .
 \label{e52}
\ee
In this expression, the symbol ``$\approx$'' comes from the fact that
all the overlapping terms in the normalization condition are assumed to
be negligible.
The prefactors $1/\sqrt{N}$ and $(2\pi\sigma_0^2)^{-1/4}$ are introduced
in order to keep $\psi_{G_1}$ normalized.
The opacities have all been set equal to one, because the probability
density associated with the plane wave (\ref{e38}) is uniform.
As it is described in more detail in Sec.~\ref{sec52}, in the passage
through $G_2$ the value of the opacity factors is not homogenous,
since the probability density associated with the wave function
reaching this grating is not uniform along $x$.
More importantly, the particular functional form displayed by
Eq.~(\ref{e52}) makes evident where nonlocality enters the problem, by
specifying the nonseparable connection between the local value of the
wave function at any point $x$ with the simultaneous action of the
partial waves coming from each slit (centered at spatially separated
points $x_j$).
This combined action has not only to do with a probability density,
but with the generation of an overall phase field that governs the
dynamical evolution of the wave function at each point of the
configuration space and can be measured through the local value of
the transverse momentum (see Sec.~\ref{sec51}).

%%%%%%%%%%%%%%%%%%%%%%%%%%%%%%%%%%%%%%%%%%%%%%%%%%%%%%%%%%%%%%%%%%%%%%%
%%%%%%%%%%%%%%%%%%%%%%%%%%%%%%%%%%%%%%%%%%%%%%%%%%%%%%%%%%%%%%%%%%%%%%%

\section{Far-field analytical results}
\label{sec4}

%%%%%%%%%%%%%%%%%%%%%%%%%%%%%%%%%%%%%%%%%%%%%%%%%%%%%%%%%%%%%%%%%%%%%%%

\subsection{Beam splitting and subsequent recombination}
\label{sec41}

\begin{figure}[t]
 \begin{center}
 \includegraphics[width=8cm]{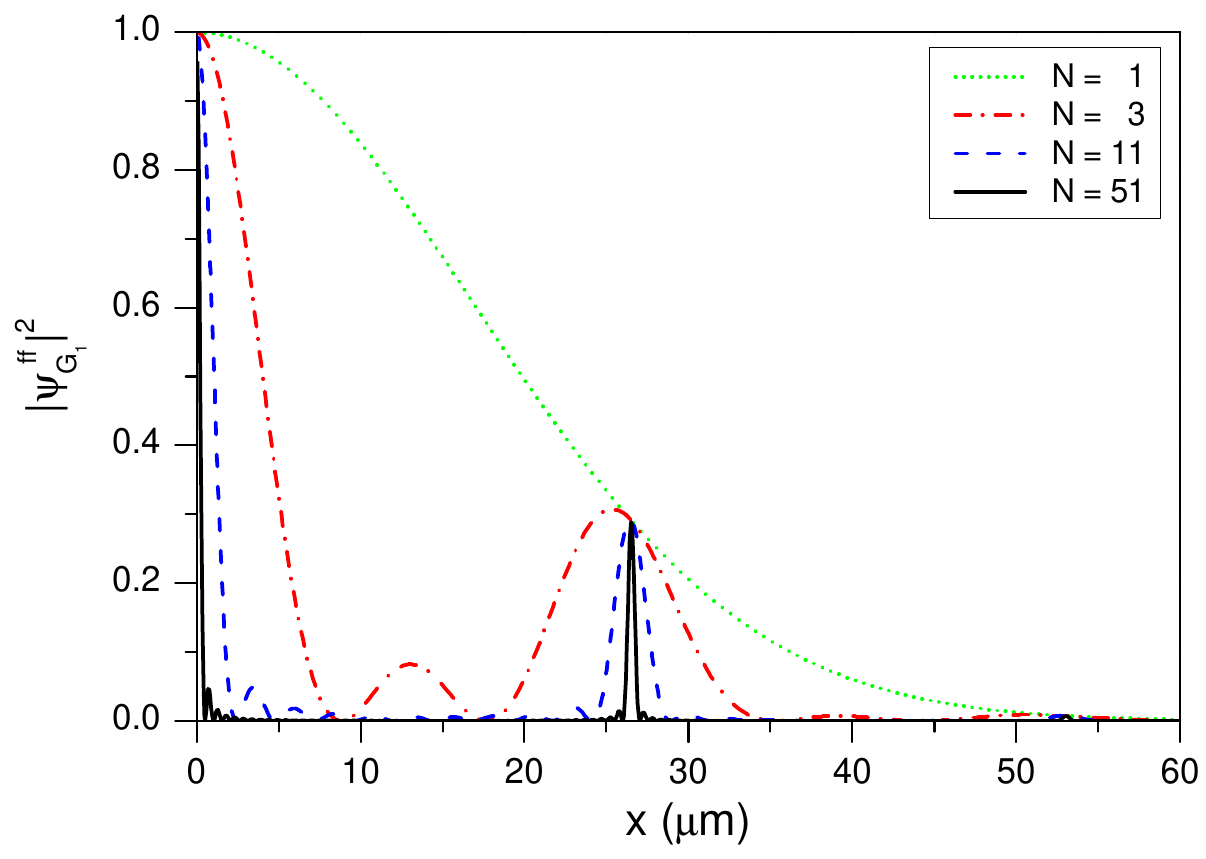}
 \caption{\label{fig2}
  Probability density $|\psi_{G_1}^{f\!f}(x,z)|^2$, given by
  Eq.~(\ref{e41m}), evaluated at $z=z_{G_2} =0.663$~m for different
  values of the number of slits (see legend); notice that the green
  dotted line corresponds to the Gaussian single-slit envelope.
  For an easier comparison, the maximum intensity has been set to one
  in all cases.}
 \end{center}
\end{figure}

The paths depicted in Fig.~\ref{fig1} behind $G_1$ or $G_2$ constitute
a convenient and simplified representation of the tracks followed by
the different diffraction orders that develop from the transmitted
atomic wave function.
To understand the origin of these paths we need to focus on how this
wave function evolves in the far field, where its shape only depends
on the aspect ratio $x/z$, as it is indicated by Eq.~(\ref{e26b}).
With this in mind, let us first consider the transit between $G_1$ and
$G_2$.
Provided the far-field condition (\ref{ffc}) is fulfilled, we can start
describing the behavior of the wave function between these two gratings
directly from Eq.~(\ref{e26b}).
If $x_j = x_0 + (j-1)d$, with $x_0 = - (N-1)d/2$ for convenience
(the origin $x_0$ is irrelevant, since it only adds a global phase
factor), we find
\be
 \psi_{G_1}^{f\!f}(x,z) \propto
  \left[ \frac{\sin (\Delta k_x Nd/2)}{\sin (\Delta k_x d/2)} \right]
  e^{-\sigma_0^2 \Delta k_x^2} ,
 \label{e41m}
\ee
where $\Delta k_x = k_x - k_{x0}$.
In this expression, the term between square brackets describes the
interference produced by the coherent superposition of the $N$
diffracted waves, while the exponential (envelope) term accounts for
the diffraction associated with a single Gaussian slit.
Equation~(\ref{e41m}) displays a series of principal maxima whenever
$k_x = k_{x,\ell} = k_{x0} + 2\pi \ell/d$ (vanishing denominator), with
$\ell = 0,\pm 1,\pm 2,\ldots$, which give rise to the corresponding
diffraction orders (other secondary maxima also appear in between,
but become meaningless as $N$ increases and are therefore physically
irrelevant).
The intensity of these diffraction orders depends on how fast the
Gaussian envelope (see the green dotted line in Fig.~\ref{fig2}) falls.
The ratio of intensities for these maxima with respect to the zeroth
diffraction order can be readily determined \cite{note1}, reading
\be
 \mathcal{R}_{0,\ell} =
 \frac{|\psi_{G_1,\ell}^{f\!f}(x,z)|^2}
      {|\psi_{G_1,0}^{f\!f}(x,z)|^2} = e^{-2(2\pi\ell\sigma_0/d)^2} .
 \label{e43c}
\ee
When numerical values are substituted into this expression, we find
that the intensity for $\ell=\pm 1$ is about 70\% weaker than for the
zeroth-order one ($\mathcal{R}_{0,\pm 1} \approx 0.29$), and that for
$\ell=\pm 2$ it is essentially negligible
($\mathcal{R}_{0,\pm 2} \approx 7.2\times 10^{-3}$).
In other words, the grating produces an effective splitting of the
amplitude of the incoming wave function into three well separated
wavefronts.
The diffraction orders $\ell=0$, $\ell=+1$, and $\ell=+2$ are displayed
in Fig.~\ref{fig2} for different values of $N$.

Given the linear relationship between $x$ and $z$ in the far field, the
position of the diffraction orders at a distance $z$ from the grating
is determined by
\be
 x_\ell = \left(\frac{k_{x,\ell}}{k} \right) z
%  = \left( \frac{k_{x0}}{k} + \frac{2\pi \ell}{kd} \right) z
  = \left( \frac{k_{x0}}{k} + \frac{\ell\lambda}{d} \right) z .
 \label{e50m}
\ee
The width of the corresponding intensity peaks, measured between the
two adjacent minima, is given by
\be
 \Delta x
%  = \left(\frac{4\pi}{kNd}\right) z
  = \left(\frac{2\lambda}{Nd}\right) z ,
 \label{e51}
\ee
which depends on $N$ (it decreases as $N$ increases), but not on the
incident wave vector, $k_{x0}$.
Substituting the numerical parameters given in Sec.~\ref{sec22} into
these expressions, we find that Eq.~(\ref{e50m}) reads as
$x_\ell = 40 \times 10^{-6}\ell z$.
That is, at a distance $z=z_{G_2}=0.663$~m from $G_1$, the two relevant
diffraction orders reach the positions
$x_{\pm 1} \approx \pm 25.42$~$\mu$m~$\approx \pm 66d$,
with a width
$\Delta x \approx 53.04 N^{-1}$~$\mu$m~$\approx 133(d/N)$.
This means that, in principle, if the far-field condition is fulfilled,
the width of the diffraction orders becomes negligible with $N$ (see
Fig.~\ref{fig2}) and their evolution along $z$ can be approximated by
the paths displayed in Fig.~\ref{fig1}.
In this figure, the two paths between $G_1$ and $G_2$ represent the
evolution along $z$ of the diffraction orders $\ell=0$ and $\ell=+1$.
Of course, only far from $G_1$ this representation in terms of
separated paths is be correct (see Sec.~\ref{sec5}).
Nonetheless, notice that because of the Gaussian envelope in
Eq.~(\ref{e41m}) it is expected a slight shift of the position described
by Eq.~(\ref{e50m}).
As seen in Fig.~\ref{fig2}, particularly for $\ell=+1$ (for $\ell=+2$
the intensity is negligible), this effect decreases as $N$ increases,
since the peak becomes narrower and narrower, eventually approaching a
$\delta$-function.

The amplitude splitting between $G_1$ and $G_2$ also takes
place between $G_2$ and $G_3$, thus closing the two ``arms'' of the
interferometer.
More specifically, what happens is that the two diffraction orders that
reach $G_2$ around $x_0$ and $x_{+1}$ give rise to two diffracted waves
around them.
Our description after $G_2$ can then be formulated in terms of a coherent
superposition of two spatially separated waves, each one leading again to
a series of diffraction orders.
Let $\ell^{(0)}$ and $\ell^{(+1)}$ be the labels for the diffraction
orders\footnote{This notation may seem to be a bit ``indigestible'', but
it is unambiguous enough to denote diffraction orders coming from
different diffracted beams generated at $G_2$.}
coming from the waves diffracted around $x_0$ and $x_{+1}$, respectively.
We notice that the diffraction orders $\ell^{(0)}=+1$ and $\ell^{(+1)}=-1$
eventually coalesce on the same spot or region on $G_3$ around
$x_{G_3} = 25.42$~$\mu$m.
As seen in Fig.~\ref{fig1}, these two paths complete the Mach-Zehnder
like structure of the atomic three-grating interferometer.

%%%%%%%%%%%%%%%%%%%%%%%%%%%%%%%%%%%%%%%%%%%%%%%%%%%%%%%%%%%%%%%%%%%%%%%

\subsection{Measurement role of the third grating}
\label{sec42}

Once the interferometer structure is demonstrated, one may wonder about
the role played by the third grating.
This grating can be laterally displaced, which allows us to modify the
flux of transmitted atoms around $x_{G_3}$, as already shown by Carnal
and Mlynek \cite{carnal1}.
Notice that unlike the maximum structures that appear around
$x = 0$~$\mu$m or $x \approx 51$~$\mu$m (the spots respectively reached
by the zeroth diffraction order of each wave), the coalescence of
$\ell^{(0)}=+1$ and $\ell^{(+1)}=-1$ produces an interference pattern
with the same period $d$ of the gratings.
This can be easily shown as follows.
The momenta of these diffraction orders are
\ba
 k_{x,\ell^{(0)} = +1} & = & k_{x,0} + 2\pi/d = 2\pi/d , \\
 k_{x,\ell^{(+1)}=-1} & = & k_{x,+1} - 2\pi/d = 0 ,
\ea
where $k_{x,0}$ and $k_{x,+1}$ are the zeroth-order momenta associated
with each one of the diffracted waves.
The corresponding momentum transfers are then equal, but opposite in
sign, i.e., $\Delta k_{x,\ell^{(0)} = +1} = k_{x,\ell^{(0)} = +1} - k_{x,0}
= 2\pi/d$ and $\Delta k_{x,\ell^{(+1)}=-1} = k_{x,\ell^{(+1)}=-1} - k_{x,+1}
= -2\pi/d$.
If Eq.~(\ref{e41m}) is evaluated replacing $G_1$ by $G_2$ (now the
starting point is $G_2$) and taking into account the above momentum
values into account, we find the far-field expression for the two
diffraction orders around $x_{G_3}$:
\ba
 \psi_{G_2,\ell^{(0)}=+1}^{f\!f}(x) & \propto &
% & \approx &
% \sqrt{\frac{2kN}{iz}} \left( \frac{\sigma_0^2}{2\pi} \right)^{1/4}
  e^{-(2\pi\sigma_0/d)^2} e^{2i\pi x/d} , \\
% \nonumber \\ & & \\
 \psi_{G_2,\ell^{(+1)}=-1}^{f\!f}(x) & \propto &
% & \approx &
% \sqrt{\frac{2kN}{iz}} \left( \frac{\sigma_0^2}{2\pi} \right)^{1/4}
  e^{-2(2\pi\sigma_0/d)^2} .
\ea
The extra factor 2 in the argument of the exponential of
$\psi_{G_2,\ell^{(+1)}=-1}^{f\!f}$ arises from the exponential prefactor
inherited from the transit between $G_1$ and $G_2$.
The coherent superposition that we find around $x_{G_3}$ is
\be
 \psi_{G_3}(x) \sim
  \psi_{G_2,\ell^{(0)}=+1}^{f\!f}(x) + \psi_{G_2,\ell^{(+1)}=-1}^{f\!f}(x)
% \nonumber \\
% & \approx & \sqrt{\frac{2kN}{iz}}
%  \left( \frac{\sigma_0^2}{2\pi} \right)^{1/4} e^{-(2\pi\sigma_0/d)^2}
%  e^{2i\pi x/d}
% \nonumber \\
% & & \qquad \qquad \times
% \left[ 1 + e^{-(2\pi\sigma_0/d)^2} e^{-2i\pi x/d} \right] ,
 \propto
 \left[ 1 + e^{-(2\pi\sigma_0/d)^2} e^{-2i\pi x/d} \right]
 e^{2i\pi x/d} ,
\ee
which gives rise to an interference pattern, described by
\ba
 |\psi_{G_3}(x)|^2
% & \propto & 1 + e^{-2(2\pi\sigma_0/d)^2}
% \nonumber \\ & & \quad
%  + 2 e^{-(2\pi\sigma_0/d)^2} \cos (2\pi x/d)
% \nonumber \\
 & \propto &
 1 + {\rm sech} [(2\pi\sigma_0/d)^2]\ \! \cos (2\pi x/d) .
 \label{probG3}
\ea
(Here we prefer the new denomination $\psi_{G_3}$ instead of
$\psi_{G_2}^{f\!f}$ because the latter refers to the full far-field
wave function reaching $G_3$ from $G_2$, while $\psi_{G_3}$ only
refers to the section of the wave function $\psi_{G_2}^{f\!f}$
around $x_{G_3}$.)
There is no full fringe visibility because each beam reaches the region
around $x_{G_3}$ with a different intensity, although the period of the
fringes is $d$, as it is for the gratings.
This is the reason why the spot around $x_{G_3}$ is the interesting one
regarding interferometry.
Because of this periodicity any perturbation happening inside the
interferometer (i.e., among the gratings) eventually translates into
a loss of fringe visibility and/or a phase shift \cite{chapman1}.

As formerly done by Carnal and Mlynek \cite{carnal1}, the number or flux
of transmitted atoms can be measured taking advantage of the
interference pattern around $x_{G_3}$.
By keeping $G_3$ aligned or misaligned with this interference pattern,
the total outgoing atomic flux collected behind this grating will be
larger or smaller, respectively.
The grating $G_3$ then acts like a mask to sample the interference
pattern.
Thus, consider that the misalignment along the $x$ direction is measured in
terms of a variable $\chi$, such that $\chi=0$ means perfect alignment
of $G_3$ with the interference pattern, and $\chi=d/2$ is maximum
misalignment.
The amount of atoms passing through $G_3$ will depend on $\chi$, and so
the total flux collected behind $G_3$.
The total flux is obtained from a convolution integral:
\be
 \mathcal{T}_{G_3}(\chi) =
  \int |\psi_{G_3}(x)|^2 T_{G_3}(\chi - x) dx
% \nonumber \\
  \propto 1 + 2 \left[1 + e^{-2(2\pi\sigma_0/d)^2}\right]^{-1}
  \cos (2\pi\chi/d) ,
 \label{flux}
\ee
where the subscript $G_3$ means that the flux is measured at $z_{G_3}$,
and $T_{G_3}$ is the transmission function associated with this
grating, which consists just of a bare sum of Gaussians [i.e., as in
Eq.~(\ref{transmit}), but with $\beta_j=0$ and $k_{j,x0}=0$ for all $j$].
Notice that although $\mathcal{T}_{G_3}$ displays a similar functional
form to the probability density (\ref{probG3}), it is a different
quantity: it measures the amount of transmitted intensity (number of
atoms) as a function of the position of the grating $G_3$ with respect
to the interference pattern (\ref{probG3}).
Nonetheless, one can measure the flux for different values of $\chi$
beyond $d$, and the measurements collected will be in direct
correspondence with the interference pattern.
An analogous functional form to Eq.~(\ref{flux}), also displaying
the same periodicity, was previously numerically
found \cite{davidovic:PhysScr:2010,davidovic:JPhysA:2012} for slit
transmission functions described by hat-functions instead of
Gaussians.
As a final remark, we would like to note that in the integral (\ref{flux})
no assumption on the spatial extension of the spot around $x_{G_3}$
has been introduced.
Obviously, the interference pattern (\ref{probG3}) has a finite
spatial extension that has to be taken into account.
In the simulations below the limits of Eq.~(\ref{flux}) have
been chosen in such a way that no contributions from nearby diffraction
orders (around $x = 0$~$\mu$m or $x \approx 51$~$\mu$m)
``contaminate'' the flux related to this interference pattern.

%%%%%%%%%%%%%%%%%%%%%%%%%%%%%%%%%%%%%%%%%%%%%%%%%%%%%%%%%%%%%%%%%%%%%%%
%%%%%%%%%%%%%%%%%%%%%%%%%%%%%%%%%%%%%%%%%%%%%%%%%%%%%%%%%%%%%%%%%%%%%%%

\section{Numerical analysis}
\label{sec5}

%%%%%%%%%%%%%%%%%%%%%%%%%%%%%%%%%%%%%%%%%%%%%%%%%%%%%%%%%%%%%%%%%%%%%%%

\subsection{Methodology}
\label{sec51}

How close are the previous analytical results to the actual evolution
of the atomic wave function inside the interferometer?
In order to investigate this question, we decided to perform a series
of numerical simulations that illustrate both the full evolution of
the wave function between gratings.
We would like to stress that the model that we are using is fully
analytical (the integration in time of each diffracted Gaussian wave
packet $\psi_j$ is
fully analytical \cite{asanz-bk2}), and therefore the numerical issue
only concerns the evaluation of the superpositions or the calculation
of some associated quantities.
Thus, it can be readily shown that the use of Eq.~(\ref{eqmrept3}), or
equivalently Eq.~(\ref{e14b}), leads to the analytical solution
\be
 \psi_{G_1}(x,z) \approx \frac{1}{\sqrt{N}}
  \left(\frac{1}{2\pi\tilde{\sigma}_z^2}\right)^{1/4}
  \sum_{j=1}^N e^{-(x - x_{j,z})^2/4\sigma_0\tilde{\sigma}_z
% \nonumber \\ & & \qquad \times
   +ik_{x0} (x-x_{j,z}) + ik_{x0} x_j + i k_{x0}^2 z/2k} ,
 \label{e52b}
\ee
where $x_{j,z} = x_j + (\hbar k_{x0}/mv)z = x_j + (k_{x0}\lambda/2\pi)z$
and $\tilde{\sigma}_z = \sigma_0 [1 + (iz/2k\sigma_0^2)]$.
In particular, for this initial state we have chosen $\beta_j=1$ for
all $j$; all the slits are assumed to be identical (see Sec.~\ref{sec22}).
This expression is very useful in the analytical derivation of some
diffraction properties in the near and far fields, as it has been done
elsewhere in more detail \cite{talbot-jcp}.

To inquire questions specifically related to the {\it local transverse
momentum} (in terms of the wave vector), we are going to introduce the
new function $K_x$, defined as
\be
 K_x(x,z) = {\rm Im} \left[\frac{1}{\psi(x,z)}
  \frac{\partial \psi(x,z)}{\partial x}\right] ,
 \label{kx}
\ee
which provides the value of the local transverse momentum $K_x$ as a
function of the $x$ coordinate at a given distance $z$ from a grating
(or whichever the $z$-axis origin is).
Notice that this momentum is connected to the usual quantum flux,
\be
 J_x = \frac{\hbar}{m}\
 {\rm Im} \left[\frac{1}{\psi} \frac{\partial \psi}{\partial x}\right]
 = \frac{\hbar}{m}\ K_x |\psi|^2 ,
\ee
commonly used in the Bohmian formulation of quantum mechanics
\cite{asanz-bk1}.
In the far field, for example, it typically coincides with the
transverse momentum value associated with the different diffraction
orders, as it can readily be seen by substituting the ansatz~(\ref{e52b})
into Eq.~(\ref{kx}) and then considering the corresponding limits
\cite{talbot-jcp}.
This renders
\be
 K_x(x,z) \approx k_{x,\ell} .
 \label{kxff}
\ee
Analogously, if we substitute the far-field expression~(\ref{e26}) into
(\ref{kx}), we find
\be
 K_x(x,z) = \frac{kx}{z} = k_x ,
\ee
which justifies our choice of $k_x$ and the recast of Eq.~(\ref{e26})
as (\ref{e26b}) in Sec.~\ref{sec22}.

%%%%%%%%%%%%%%%%%%%%%%%%%%%%%%%%%%%%%%%%%%%%%%%%%%%%%%%%%%%%%%%%%%%%%%%
%%%%%%%%%%%%%%%%%%%%%%%%%%%%%%%%%%%%%%%%%%%%%%%%%%%%%%%%%%%%%%%%%%%%%%%

\subsection{Results}
\label{sec52}

\begin{figure}[t]
 \begin{center}
 \includegraphics[width=16cm]{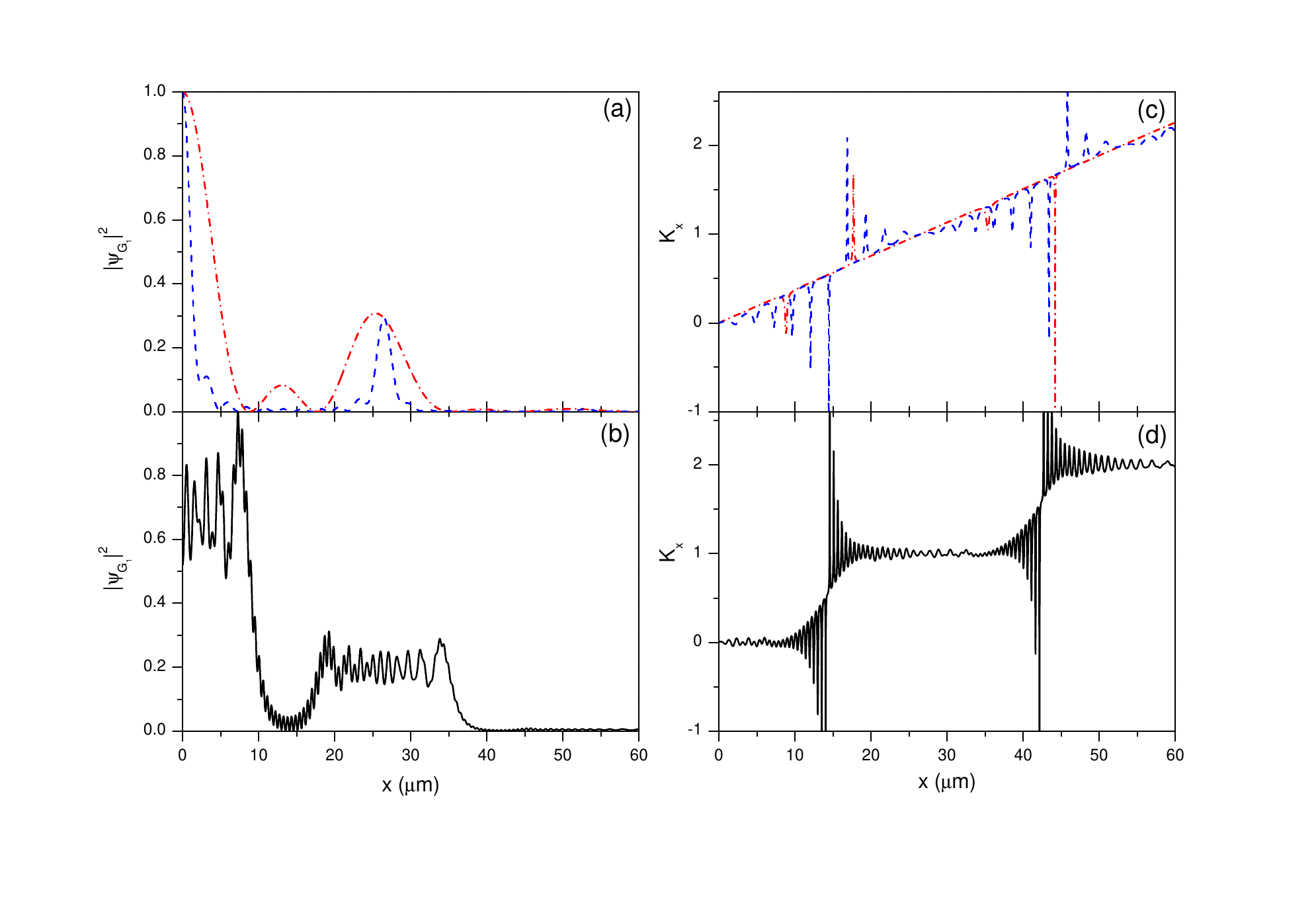}
 \caption{\label{fig3}
  Probability density $|\psi_{G_1}(x,z)|^2$ (left column) and local
  transverse momentum $K_x$ (right column) evaluated at the position
  of the grating $G_2$ ($z=z_{G_2}=0.663$~m) for the same number of
  slits considered in Fig.~\ref{fig2} and numerically propagated
  according to Eq.~(\ref{e52b}).
  For visual clarity, panels (a) and (c) are for $N=3$ (red dash-dotted
  line) and $N=11$ (blue dashed line), and panels (b) and (d) for $N=51$.
  For an easier comparison, in panels (a) and (b) the maximum intensity
  has been set to one in all cases, while in panels (c) and (d) the
  momentum is given in units of $2\pi/d$.}
 \end{center}
\end{figure}

In order to show the reliability on the far-field expression given by
Eq.~(\ref{e41m}) at $z=z_{G_2}=0.663$~$\mu$m, where the grating $G_2$
is placed, the probability density $|\psi_{G_1}(x,z_{G_2})|^2$ is
plotted in Fig.~\ref{fig3}(a) for $N=3$ (red dashed-dotted line) and
$N=11$ (blue dashed line), and in Fig.~\ref{fig3}(b) for $N=51$
(we have split up the graph for visual clarity).
When comparing the results of this figure with the homologous cases in
Fig.~\ref{fig2}, we find that the agreement gets worse as $N$ increases,
which can already be noticed in the nonzero secondary minima
observed in Fig.~\ref{fig3}(a).
This effect is related to the fact that, as $N$ increases, the
near-field region spreads further away and therefore longer distances
from the grating than $z_{G_2}$ need to be considered in order to
satisfy the far-field condition.
Actually, as seen in panel (b), for even larger $N$ the expected
very narrow diffraction orders do not appear, but rather wide
diffraction, plateau-like structures, with a width analogous to
the one experimentally reported of about 30~$\mu$m \cite{keith2}.
In this latter case, although there are well separated diffraction
orders, they are not of the kind described by the far-field expression
given by Eq.~(\ref{e41m}).
This expression was obtained under the assumption
that the wave is already in the far field, where the number of slits
only influences the width and number of maxima ---this approximated
expression rules the behavior of the wave function along the
transverse direction without taking into account the longitudinal one.

We are very familiar with the probability density, but what about the
transverse momentum?
Does this momentum agree with the assumption that it should be equal
to the one carried by each diffraction order?
This issue can be easily analyzed by inspecting the right-hand side
panels of Fig.~\ref{fig3}, where the transverse momentum
$K_x(x,z_{G_2})$ is shown for $N=3$ (red dash-dotted line) and $N=11$
(blue dashed line) in panel (c), and for $N=51$ in panel (d) (again,
the graph has been split up for visual clarity).
Surprisingly, as seen in panel (c), for low $N$ we find no trace of the
momenta related to the diffraction orders, $k_{x,\ell}$.
In these cases, $K_x$ essentially behaves linearly with the $x$
coordinate in compliance with the relation $K_x = (k/z)x$, except at
some particular values, where a kind of sudden positive or negative
kink is observed.
The $x$ values at which this behavior appears are those for which
$|\psi_{G_1}(x,z_{G_2})|^2$ vanishes (nodes between maxima).
As can be seen, all the spikes that appear in the region from the
center of a diffraction order to half the distance
between this order and the next one are below the line described by
the relation $K_x = (k/z)x$; all the spikes in the opposite region are
above this line.
This indicates a trend: the quantum flux tends to be redirected along
the directions of the diffraction orders.
Close to the position of the diffraction order, the spikes are
relatively weak, while as we move far from it they start increasing.
To the left of the diffraction order they are positive, which causes
a net effect of pushing the quantum flux towards the right, i.e.,
approaching it to the position of the diffraction order.
On the other hand, as we move to the right of the diffraction order,
the spikes increase negatively, pushing the flow leftwards.
The combination of both effects leads to an effective confinement
or ``quantization'' of the quantum flux around the corresponding
diffraction orders.
This becomes more apparent as $N$ gets significantly larger [see panel
(d)], when $K_x$ starts displaying a staircase structure.
This structure corresponds to the momentum
quantization associated with the appearance of separated diffraction
orders; the emerging well-defined values or steps of (transverse)
momenta are precisely $k_{x,\ell}$.
This staircase allows us to specifically determine the
spatial domain associated with each diffraction order, which has some
computational advantages with respect to the numerical simulation of
the wave function evolution behind $G_2$, as seen below.

\begin{figure}[t]
 \begin{center}
 \includegraphics[width=16cm]{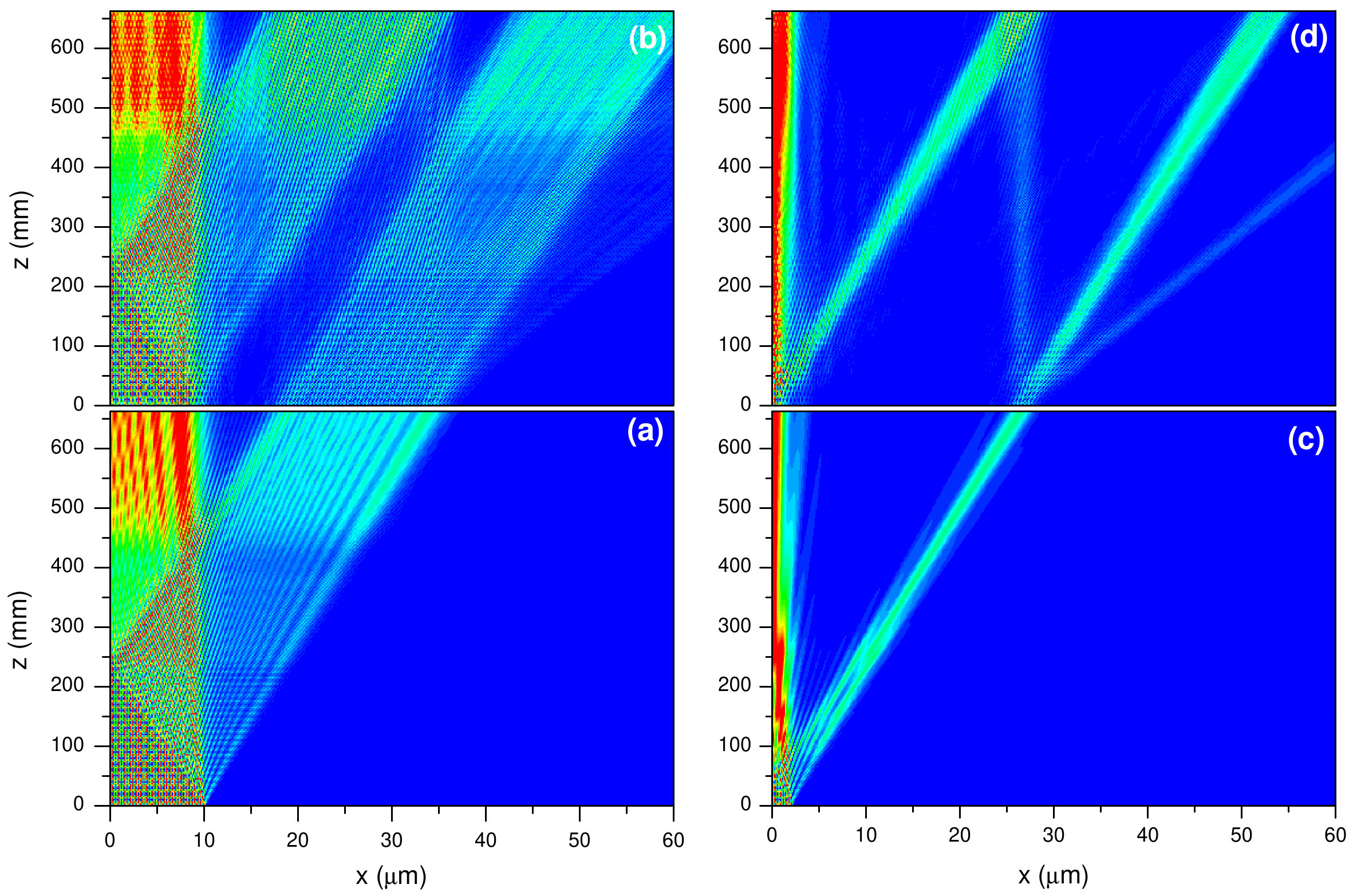}
 \caption{\label{fig4}
  Contour-plots of the probability density between $G_1$ and $G_3$ for
  $N=51$ (left) and $N=11$ (right).
  The evolution of $|\psi_{G_1}(x,z)|^2$ is displayed in panels (a) and
  (c) $|\psi_{G_1}(x,z)|^2$, while $|\psi_{G_2}(x,z)|^2$ is given in
  panels (b) and (d).
  At every value of the $z$-coordinate, the maximum value of the
  probability density has been set to one; a truncation at 0.8 has then
  been considered for visual clarity.}
 \end{center}
\end{figure}

The full evolution of the probability density between $G_1$ and $G_3$
is displayed in Fig.~\ref{fig4} for $N=51$ (left) and $N=11$ (right),
respectively, in the form of contour-plots [color scale: maxima are
denoted by red, while minima are indicated by blue; because of the
faint visibility of the diffraction order $\ell=-1$ in the upper
panels (a) and (c), we have introduced a truncation of the maximum
contour].
The propagation from $G_1$ to $G_2$ [see panels (b) and (d)] has been
carried out considering that all the slits illuminated contribute with
the same weight, while in the transit from $G_2$ to $G_3$
[see panels (a) and (c)] we have considered $\beta_j$ to be proportional
to the square root of the probability density reaching $G_2$ at the
center of the corresponding slit, $j$.
In this latter case, if the probability density (normalizing its
maximum to unity) was smaller than 0.01, the corresponding $\beta_j$
was chosen to be zero.
Regarding the momenta $k_{j,x0}$ for the diffracted wave beyond $G_2$,
we have considered the domains of Fig.~\ref{fig3}(d), setting
$k_{j,x0} = k_{x,\ell}$ for all $j$ such that $x_{j,0}$ is confined
within the region associated with the $\ell$th diffraction order.
Given the limited number of Gaussians used to simulate the transit
from $G_2$ to $G_3$ according to the above prescription (329 for $N=51$
and 265 for $N=11$), it has been observed that a direct identification
of $k_{j,x0}$ with the local value of $K_x$ at $x_{0,j}$ introduces
remarkable numerical errors into the simulations due to the fast
oscillatory behavior of $K_x$ as $N$ increases.
This is the reason why this second method to choose $k_{j,x0}$ has been
rejected in the current work ---although a priori it may seem physically
reasonable, it carries computational disadvantages.

\begin{figure}[t]
 \begin{center}
 \includegraphics[width=16cm]{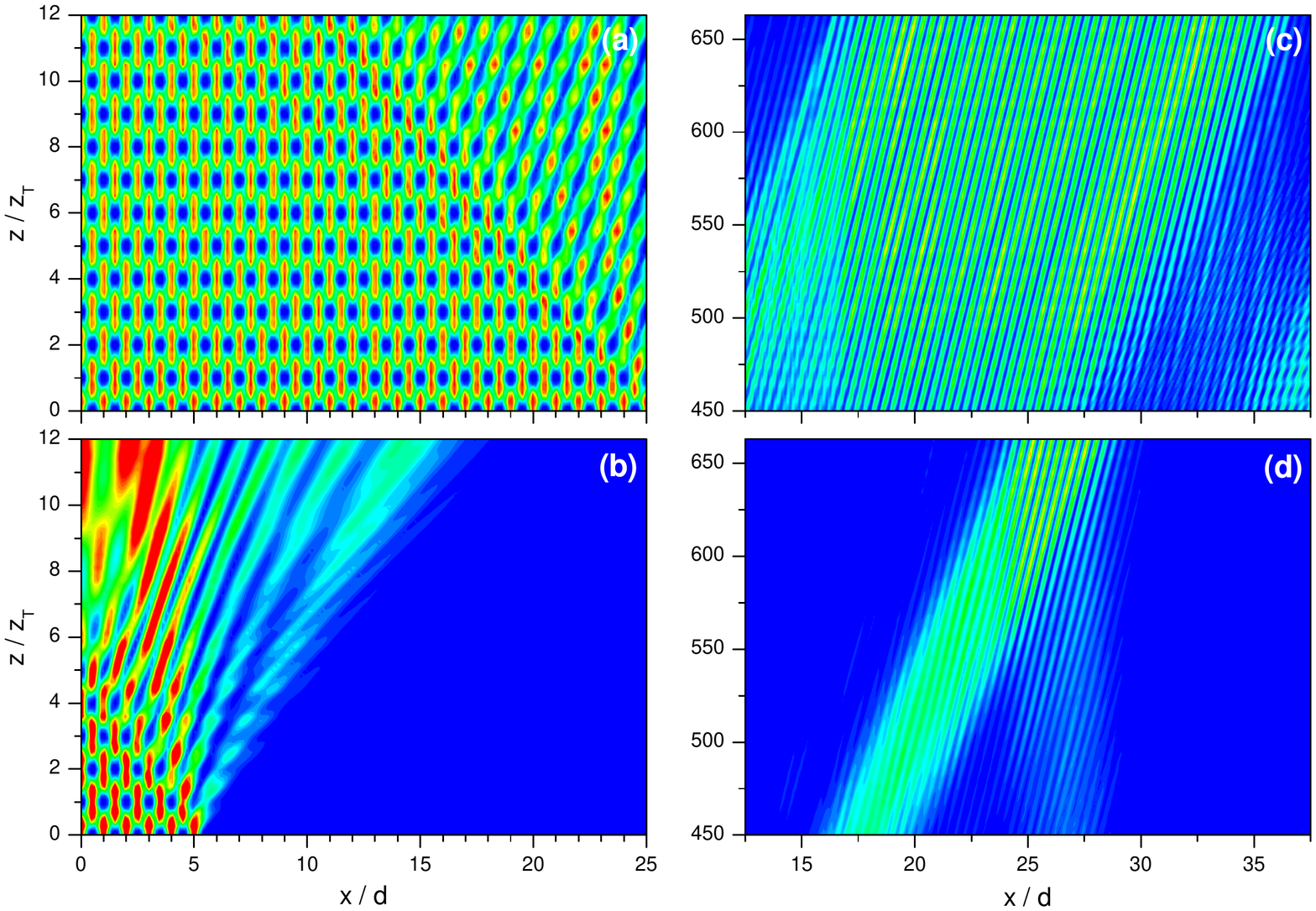}
 \caption{\label{fig5}
  Magnifications to emphasize a series of aspects from Fig.~\ref{fig4}.
  Left: Contour-plots of the probability density $|\psi_{G_1}(x,z)|^2$
  in the near field (close to $G_1$) showing the emergence of the Talbot
  carpet for $N=51$ (a) and $N=11$ (b).
  Right: Contour-plots of the probability density $|\psi_{G_2}(x,z)|^2$
  in the far field (close to $G_3$) for for $N=51$ (c) and $N=11$ (d),
  where interference features due to the coalescence of the $\ell_0=+1$
  and $\ell_{+1}=-1$ diffraction orders coalesce around $x = 25.42$~$\mu$m
  are apparent.
  At every value of the $z$-coordinate, the maximum value of the
  probability density has been set to one for visual clarity.}
 \end{center}
\end{figure}

The two cases considered in Fig.~\ref{fig4} demonstrate the Mach-Zehnder
interferometer configuration formed by the corresponding diffraction
orders.
Because the diffraction orders are narrower in the case $N=11$, a
certain widening is observed from $G_2$ to $G_3$, as $z$ increases
---this behavior is expected from relatively narrow wave packets
\cite{sanz-jpa}.
Nonetheless, the most relevant aspect, common to both cases, is the
relatively complex structure displayed by the wave function along its
evolution.
Although at a qualitative level one can represent the interferometer
in terms of the paths displayed in Fig.~\ref{fig1} (i.e., particle-like
behavior), a closer inspection reveals a rather convoluted structure
due to the wave nature of the atomic beam all the way through, which
cannot be just neglected.
In this regard, two structures are worth discussing, namely the
near-field carpet that can be seen just behind each grating, and the
interference pattern in the far field observable in panels (b) and (d).

The repetitive structures near both $G_1$ and $G_2$ are associated with
a typical effect of periodic gratings on extended waves, namely the
so-called Talbot effect
\cite{talbot,rayleigh,winthrop,latimer,talbot-jcp,davidovic-talbot}.
This effect is shown with more detail in the enlargements near $G_1$
displayed in the left-hand side panels of Fig.~\ref{fig5}.
The grating acts on the impinging wave as a collimator that generates
a series of identically diffracted waves.
The periodicity in the distribution of these waves is such that, if one
recast each diffracted wave as a superposition of plane waves, a certain
quantization condition arises, which only allows certain
momenta \cite{talbot-jcp}.
The larger the number of slits the lesser the number of allowed
momenta, until reaching a minimum given by the ideal case of
$N\to\infty$.
Due to this finite basis of momenta, as $z$ increases we observe that
the probability density displays recurrences along $x$ within a
distance $d$.
Thus, at distances $z = 2z_T = 2d^2/\lambda = 20$~mm we find an exact
copy of the initial probability density, where $z_T$ is the so-called
Talbot distance.
Now, because there are no physical boundaries separating different
slits, we also find exact copies of the probability density at the
Talbot distance, although they have a half a period displacement with
respect to the initial pattern.
These copies will cover a larger spatial region as $N$ increases,
because the basis of momenta will be more limited (compare panels (a)
and (b)).
Also, if the transmission is different for each slit, we can still
observe the Talbot carpet, although with some distortions (see
Figs.~\ref{fig4}(b) and (d) near $G_2$).

\begin{figure}[t]
 \begin{center}
 \includegraphics[width=16cm]{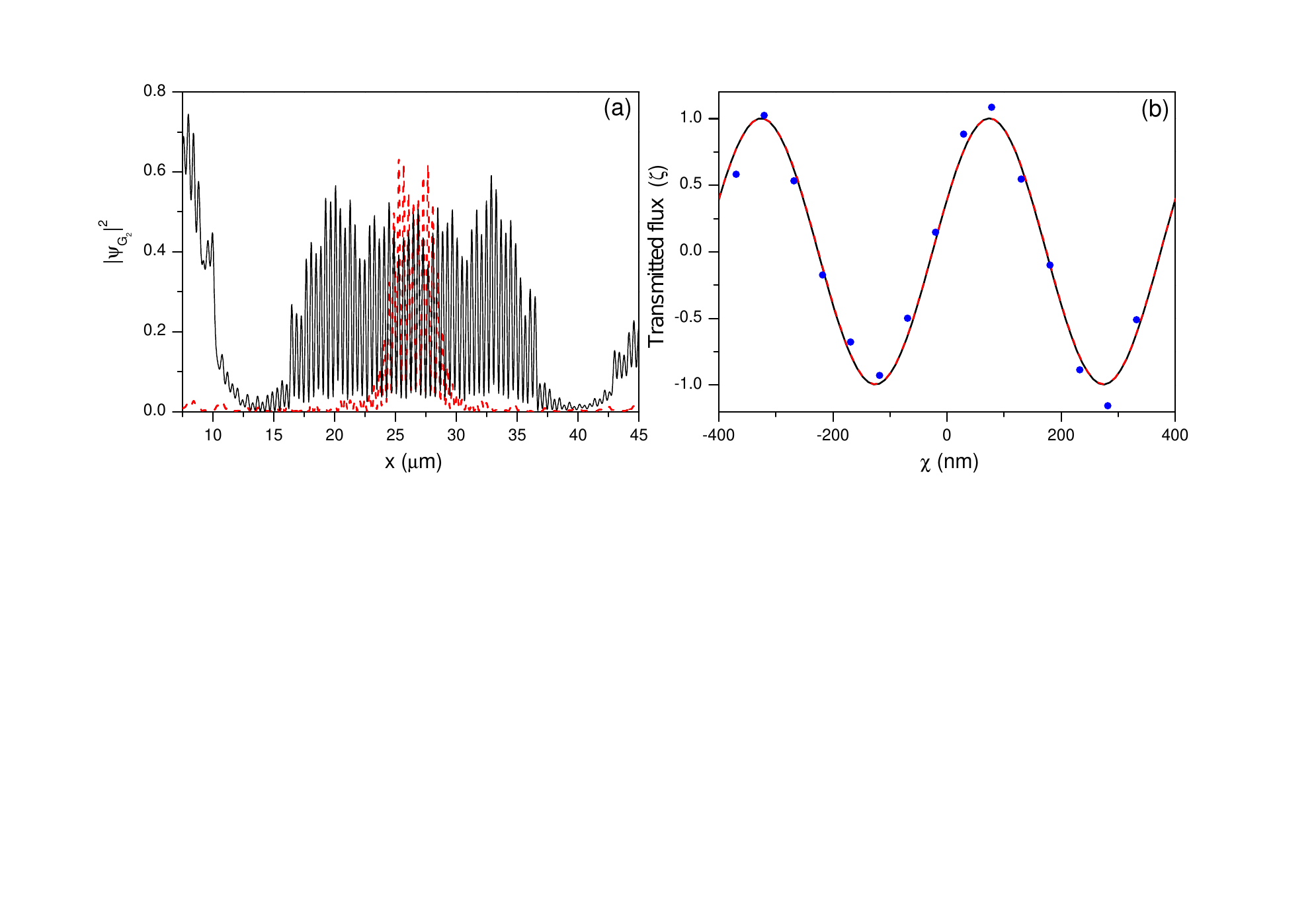}
 \caption{\label{fig6}
  (a) Probability density $|\psi_{G_2}(x,z_{G_2})|^2$ at $G_3$ for
  $N=51$ (black solid line) and $N=11$ (red dashed line).
  (b) Convolution of the grating transmission function with the
  intensity around $x_{G_3}$, according to Eq.~(\ref{flux}).
  To compare with, the experimental data provided in Ref.~\cite{keith2}
  are displayed as full (blue) circles.
  In all cases the transmitted flux is expressed in terms of the
  normalized function $\zeta$ (see text for details).}
 \end{center}
\end{figure}

In the far field with respect to $G_2$, just around $x_{G_3}$ at $G_3$,
we find the the second type of pattern, as can be seen in the right-hand
side panels of Fig.~\ref{fig5}.
Regardless of the number of illuminated slits, this pattern has a period
$d$.
It is then clear that if $G_3$ is gradually moved laterally with respect
to this pattern, for some positions the atomic flux passing through the
grating will be maximum (in phase), and for other it will be minimum
(out of phase), observing a periodic variation (with period $d$), in
compliance with Eq.~(\ref{flux}).
As it was stressed above, the emergence of this pattern is precisely
the reason why this peak is important for interferometry: any small
perturbation on any of the two paths selected between $G_1$ and $G_2$
affects its fringe visibility, which can be directly detected through
the amount of flux measured behind $G_3$.
The interference pattern at $G_3$ and the amount of transmitted flux as
a function of the displacement $\chi$ for $N=51$ and $N=11$ are plotted
in Figs.~\ref{fig6}(a) and (b), respectively, where the experimental
data reported in Ref.~\cite{keith2} are also included (full blue circles
in panel (b)).
More specifically, in order to avoid contamination from other adjacent
diffraction structures, we have performed the convolution integral in
both cases within the range between $x=20$~$\mu$m and $x=32.5$~$\mu$m
(just around $x_{G_3}$), with $\chi$ varying between $-6d$ and $6d$
---in the figure only two periods are shown.
In agreement with Eq.~(\ref{flux}), the results of panel (b) display
the same cosine dependence with $\chi$ in either case, in good
agreement with the experimental data reported.
Nonetheless, because the total amount of flux collected is larger
for $N=51$ than for $N=11$ (the extension covered by the interference
pattern is larger), and also for a better comparison with the
experimental data, instead of directly plotting Eq.~(\ref{flux}),
we have considered the normalized function
\be
 \zeta = \frac{\mathcal{T}_{G_3}(\chi) - \bar{\mathcal{T}}_{G_3}}
   {\Delta \mathcal{T}_{G_3}} ,
\ee
where $\bar{\mathcal{T}}_{G_3} = (\mathcal{T}_{G_3}^{\rm max}
+ \mathcal{T}_{G_3}^{\rm min})/2$ and
$\Delta \mathcal{T}_{G_3} = (\mathcal{T}_{G_3}^{\rm max}
- \mathcal{T}_{G_3}^{\rm min})/2$.
Moreover, the maxima of the numerical calculations have also been
shifted in order to align them with those of the experimental data.
Nonetheless, unlike the models considered in the literature, we would
like to highlight that here no best-fit procedure has been used in
any of the steps to adjust our results to the reported experimental
data; the agreement between our simulations and the experiment
directly arises from the few working hypotheses considered above.

%%%%%%%%%%%%%%%%%%%%%%%%%%%%%%%%%%%%%%%%%%%%%%%%%%%%%%%%%%%%%%%%%%%%%%%
%%%%%%%%%%%%%%%%%%%%%%%%%%%%%%%%%%%%%%%%%%%%%%%%%%%%%%%%%%%%%%%%%%%%%%%

\section{Concluding remarks}
\label{sec6}

We have shown that by using the relationship between the configuration
and momentum representations of a wave function it is possible a simple
analysis of atomic three-grating Mach-Zehnder interferometers.
This analysis shows how three gratings behave in the same way as the
set of two beam-splitters and two mirrors does in conventional optical
Mach-Zehnder interferometers.
As a convenient working model we have considered a Gaussian grating
given its analytical and computational advantages.
On the one hand, this type of grating has allowed
us to obtain a series of analytical expansions and results in the far
field, which can be readily compared to the experiment, providing us
with an important physical insight on the interferometers here considered.
On the other hand, the fact that the time-evolution of Gaussian wave packets
(or, equivalently, their propagation along the longitudinal direction) is
fully analytical, also constitutes a remarkable simplification in the
design of simple numerical codes that allow to compute the probability
density or the local transverse momentum at any intermediate step between
two consecutive gratings.
In this regard, the local transverse momentum has been introduced as an important
analytical tool, borrowed from the Bohmian formulation of quantum mechanics
and related to the usual flux operator.
Based on this working model, a reasonable ansatz for the passage through
the second grating can also be easily proposed, without any need to use
more complex and higher time-consuming wave-packet propagation techniques
\cite{zhang-bk},
where the presence of gratings is usually introduced in terms of
interaction potentials \cite{sanz-ch-micha}.

The information obtained from the analytical and numerical results not
only complements each other, but is very valuable to understand different
aspects of how atomic three-grating Mach-Zehnder interferometers work.
From experimental data, the analytical results have shown in a simple
manner how the two branches of the interferometer appear as well as how,
later on, they coalesce and give rise to an interference structure with
the same period of the gratings.
In particular, first one readily notices that because the diffraction
orders are well separated spatially, one can work with only two of
them, neglecting other contributions.
This fact relies on the property that only the probability
along the diffraction orders $\ell = 0$ and $\pm 1$ is physically
relevant.
This has been confirmed by the numerical simulations.
In spite of the rather complex evolution displayed by the wave function
between consecutive gratings, particularly in the corresponding near
field regions, in practice this does not count much, this showing
the correctness of oversimplified sketches like the one displayed in
Fig.~\ref{fig1}.
Hence, bringing back the old wave-corpuscle dichotomy, we have shown
that wave and particle are not incompatible aspects, but the misuse
that we typically do of them to explain and understand these
experiments.
The incident atomic beam behaves as a wave all the way through, although
in the transit one can simplify the picture by assuming that the atoms
behave as classical point-like particles moving along straight lines.
Of course, this is done at the expense of neglecting the rich
interference structures that arise along the way.

From a practical viewpoint, we would also like to highlight the fact
that the simplicity of the model here considered makes feasible the
analysis of incoherence due to a lack of full periodicity in the
gratings, presence of thermal vibrations, or decoherence by an
external environment in a simple manner.
This can be done by playing around with the different parameters
associated with the Gaussians as well as with the way how the latter
overlap.
Indeed, both the model and the methodology here developed are not
constrained to the system that we have analyzed, but they can be easily
and conveniently implemented in other experimental contexts due to
their versatility.
Notice that, generally speaking, the main purpose of this work
consists in providing a clear, precise and simple methodology (working
model) to simulate, analyze, understand and explain interference
processes and interferometry experiments.
This is something in the borderline between the simplistic approaches
often considered in the literature (which always require of fitting
parameters and do not account for the full dynamical process that takes
place inside the interferometer), and tough and serious (very realistic)
quantum-dynamical calculations implying determining the potential
energy surfaces associated with the interaction between the diffracted
particles and the diffracting gratings as a function of the distance
(e.g., atom-atom, atom-molecule or molecule-molecule scattering
processes).
In a few words, by means of this procedure we have tried to
{\it make quantum mechanics less mystic}, showing that a relatively
complex dynamical process can be easily explained by means of a few
working hypotheses and a simple model.

%%%%%%%%%%%%%%%%%%%%%%%%%%%%%%%%%%%%%%%%%%%%%%%%%%%%%%%%%%%%%%%%%%%%%%%
%%%%%%%%%%%%%%%%%%%%%%%%%%%%%%%%%%%%%%%%%%%%%%%%%%%%%%%%%%%%%%%%%%%%%%%

\section*{Acknowledgements}

Support from the Ministerio de Econom\'{\i}a y Competitividad (Spain)
under Project No.\ FIS2011-29596-C02-01 (AS) as well as a ``Ram\'on y
Cajal'' Research Fellowship with Ref.\ RYC-2010-05768 (AS), and the
Ministry of Education, Science and Technological Development of Serbia
under Projects Nos.\ OI171005 (MB), OI171028 (MD), and III45016
(MB, MD) is acknowledged.

%%%%%%%%%%%%%%%%%%%%%%%%%%%%%%%%%%%%%%%%%%%%%%%%%%%%%%%%%%%%%%%%%%%%%%%
%%%%%%%%%%%%%%%%%%%%%%%%%%%%%%%%%%%%%%%%%%%%%%%%%%%%%%%%%%%%%%%%%%%%%%%

%\bibliography{references}

\begin{thebibliography}{55}
\providecommand{\natexlab}[1]{#1}
\providecommand{\url}[1]{\texttt{#1}}
\providecommand{\urlprefix}{URL }
\expandafter\ifx\csname urlstyle\endcsname\relax
  \providecommand{\doi}[1]{doi:\discretionary{}{}{}#1}\else
  \providecommand{\doi}[1]{doi:\discretionary{}{}{}\begingroup
  \urlstyle{rm}\url{#1}\endgroup}\fi
\providecommand{\bibinfo}[2]{#2}

\bibitem{badurek}
 G. Badurek, H. Rauch, A. Zeilinger (Eds.), Proceedings of the
 International Workshop on Matter Wave Interferometry
 (Vienna 1987), Physica B 151 (1988) 3-400.

\bibitem{adams}
 C.S. Adams, M. Sigel, J. Mlynek, Atom optics,
 Phys. Rep. 240 (1994) 143--210.

\bibitem{berman}
 P. Berman, Atom Interferometry, Academic Press, San Diego, 1997.

\bibitem{cronin}
 A.D. Cronin, J. Schmiedmayer, D.E. Pritchard, Optics and interferometry
 with atoms and molecules, Rev. Mod. Phys. 81 (2009) 1051--1129.

\bibitem{arndt1}
 M. Arndt, K. Hornberger, Quantum Interferometry with Complex Molecules,
 in Proceedings of the International School of Physics ``Enrico Fermi'',
 Course CLXXI ``Quantum Coherence in Solid State Systems'',
 P. Schwendimann (Ed.), Societ\'a Italiana di Fisica, 2008.

\bibitem{arndt5}
 M. Arndt, A. Ekers, W. von Klitzing, H. Ulbricht, Focus on
 modern frontiers of matte wave optics and interferometry,
 New J. Phys. 14 (2012) 125006(1-9).

\bibitem{kasevich:PRL:1991-2}
 M. Kasevich, S. Chu, Atomic interferometry using stimulated
 Raman transitions, Phys. Rev. Lett. 67 (1991) 181--184.

\bibitem{stern}
 I. Estermann, O. Stern, Beugung von Molekularstrahlen,
 Z. Physik 61 (1930) 95--125.

\bibitem{gould}
 P.L. Gould, G.A. Ruff, D.E. Pritchard, Diffraction of atoms by light:
 The near-resonant Kapitza-Dirac effect,
 Phys. Rev. Lett. 56 (1986) 827--830.

\bibitem{kapitza}
 P.L. Kapitza, P.A.M. Dirac, The reflection of electrons from standing
 light waves, Proc. Cambridge Phil. Soc. 29 (1933) 297--300.

\bibitem{batelaan}
 D.L. Freimund, K. Aflatooni, H. Batelaan, Observation of the
 Kapitza-Dirac effect, Nature 413 (2001) 142--143.

\bibitem{keith1}
 D.W. Keith, M.L. Schattenburg, H.I. Smith, D.E. Pritchard, Diffraction
 of atoms by a transmission grating, Phys. Rev. Lett. 61 (1988) 1580--1583.

\bibitem{keith2}
 D.W. Keith, C.R. Ekstrom, Q.A. Turchette, D.E. Pritchard, An
 interferometer for atoms, Phys. Rev. Lett. 66 (1991) 2693--2696.

\bibitem{rasel}
 E.M. Rasel, M.K. Oberthaler, H. Batelaan, J. Schmiedmayer, A. Zeilinger,
 Atom wave interferometry with diffraction gratings of light,
 Phys. Rev. Lett. 75 (1995) 2633--2637.

\bibitem{carnal2}
 O. Carnal, A. Faulstich, J. Mlynek, Diffraction of metastable helium atoms
 by a transmission grating, Appl. Phys. B 53 (1991) 88--91.

\bibitem{zeilinger3}
 B. Brezger, L. Hackerm\"uller, S. Uttenthaler, J. Petschinka, M. Arndt,
 A. Zeilinger, Matter-wave interferometer for large molecules,
 Phys. Rev. Lett. 88 (2002) 100404(1--4).

\bibitem{arndt3}
 K. Hornberger, S. Gerlich, P. Haslinger, S. Nimmrichter, M. Arndt,
 Colloquium: Quantum interference of clusters and molecules,
 Rev. Mod. Phys. 84 (2012) 157--173.

\bibitem{arndt4}
 T. Juffmann, A. Milic, M. M\"ullneritsch, P. Asenbaum, A. Tsukernik,
 J. T\"uxen, M. Mayor, O. Cheshnovsky, M. Arndt, Real-time single-molecule
 imaging of quantum interference, Nat. Nanotech. 7 (2012) 297--300.

\bibitem{clauser1}
 J.F. Clauser, S. Li, Talbot-vonLau atom interferometry with cold slow
 potassium, Phys. Rev. A 49 (1994) R2213--R2216.

\bibitem{clauser2}
 J.F. Clauser, M.W. Reinsch, New theoretical and experimental results in
 Fresnel optics with applications to matter-wave and X-ray interferometry,
 Appl. Phys. B 54 (1992) 380--395.

\bibitem{talbot}
 H.F. Talbot, Facts relating to optical science,
 Phil. Mag. 9 (1836) 401--407.

\bibitem{rayleigh}
 L. Rayleigh, On copying diffraction-gratings, and some phenomena
 connected therewith, Philos. Mag. 11 (1881) 196--205.

\bibitem{winthrop}
 J.T. Winthrop, C.R. Worthington, Theory of Fresnel Images. I. Plane
 periodic objects in monochromatic light.
 J. Opt. Soc. Am. 55 (1965) 373--381.

\bibitem{latimer}
 P. Latimer, R.F. Crouse, Talbot effect reinterpreted,
 Appl. Opt. 31 (1992) 80--89.

\bibitem{lau}
 E. Lau, Beugungserscheinungen an Doppelrastern,
 Ann. Phys. 6 (1948) 417--423.

\bibitem{lohmann}
 J. Jahns, A.W. Lohmann, The Lau effect: A diffraction experiment with
 incoherent illumination, Opt. Commun. 28 (1979) 263--267.

\bibitem{sudol}
 R. Sudol, B.J. Thompson, Lau effect: Theory and experiment,
 Appl. Opt. 20 (1981) 1107--1116.

\bibitem{storey:JPIIF:1994}
 P. Storey, C. Cohen-Tannoudji, The Feynman path integral approach to
 atomic interferometry. A tutorial, J. Phys. II France 4 (1994) 1999--2027.

\bibitem{nachman}
 P. Nachman, Mach-Zehnder interferometry as an instructional tool,
 Am. J. Phys. 63 (1995) 39--43.

\bibitem{asanz-bk1}
 A.S. Sanz, S. Miret-Art\'es, A Trajectory Description of Quantum
 Processes. I. Fundamentals, Springer, Berlin, 2012.

\bibitem{schiff-bk}
 L.I. Schiff, Quantum Mechanics, McGraw-Hill, Singapore, 1968.
% p. 22-27

\bibitem{zhang-bk}
 J.Z.H. Zhang, Theory and Application of Quantum Molecular Dynamics,
 World Scientific, Singapore, 1999.

\bibitem{sanz-jpa}
 A.S. Sanz, S. Miret-Art\'es, A trajectory-based understanding of quantum
 interference, J. Phys. A 41 (2008) 435303(1-23).

\bibitem{sanz-ch-micha}
 A.S. Sanz, S. Miret-Art\'es, Atom-surface diffraction: A quantum
 trajectory description, in Quantum Dynamics of Complex Molecular Systems,
 D.A. Micha and I. Burghardt (Eds.), Springer, Berlin, 2006, pp.~343--368.

\bibitem{miller:JCP-2:2001}
 R. Gelabert, X. Gim\'enez, M. Thoss, H. Wang, W. H. Miller,
 Semiclassical description of diffraction and its quenching by the
 forward-backward version of the initial value representation,
 J. Chem. Phys. 114 (2001) 2572--2579.

\bibitem{grisenti1}
 R.E. Grisenti, W. Sch\"ollkopf, J.P. Toennies, G.C. Hegerfeldt,
 T. K\"ohler, Determination of atom-surface van der Waals potential
 from transmission-grating diffraction intensities,
 Phys. Rev. Lett. 83 (1999) 1755--1758.

\bibitem{grisenti2}
 R.E. Grisenti, W. Sch\"ollkopf, J.P. Toennies, J.R. Manson, T.A. Savas,
 H.I. Smith, He-atom diffraction from nanostructure transmission gratings:
 The role of imperfections, Phys. Rev. A 61 (2000) 033608(1--15).

\bibitem{feynman-hibbs}
 R. P. Feynman and A. R. Hibbs, Quantum Mechanics and Path Integrals,
 McGraw-Hill, New York, 1965.
 Emended version: R. P. Feynman, A. R. Hibbs, D. F. Styer, Quantum Mechanics
 and Path Integrals, Dover, New York, 2010.

\bibitem{davidovic:PhysScr:2010}
 M. Bo\v{z}i\'c, D. Arsenovi\'c, A.S. Sanz, M. Davidovi\'c,
 On the influence of resonance photon scattering on atom interference,
 Phys. Scr. T140 (2010) 014017(1--5).

\bibitem{ballentine-bk}
 L.E. Ballentine, Quantum Mechanics: A Modern Development,
 World Scientific, Singapore, 1998.

\bibitem{dimic}
 M. Bo\v zi\'c, D. Dimi\'c, M. Davidovi\'c, Coherent beam splitting
 by a thin grating, Acta Phys. Pol. A 116 (2009) 479--482.

\bibitem{lipson3-bk}
 A. Lipson, S.G. Lipson, H. Lipson, Optical Physics,
 Cambridge University Press, Cambridge, 2011, 4th Ed.

\bibitem{hecht-bk}
 E. Hecht, Optics, Addison-Wesley Longman, New York, 2002, 4th Ed.

\bibitem{sanz2}
 A.S. Sanz, F. Borondo, and S. Miret-Art\'es, ``Particle diffraction
 studied using quantum trajectories,'' J. Phys.: Condens. Matter
 {\bf 14}, 6109-6145 (2002).

\bibitem{note1}
 This can be easily shown by taking into account the relations:
 \ba
  \sin(2N\theta) & = & 2\sin(N\theta)\cos(N\theta), \nonumber \\
  \sin(N\theta) & = & \sum_{n=0}^N
   \left(\begin{array}{c} N \\ n \end{array}\right)
   (\sin \theta)^{N-n} (\cos \theta)^n \sin[(N-n)\pi/2] , \nonumber \\
  \cos(N\theta) & = & \sum_{n=0}^N
   \left(\begin{array}{c} N \\ n \end{array}\right)
   (\sin \theta)^{N-n} (\cos \theta)^n \cos[(N-n)\pi/2] , \nonumber
 \ea
 with $\theta = \Delta k_x d/4$.

\bibitem{carnal1}
 O. Carnal, J. Mlynek, Young's double-slit experiment with atoms:
 A simple atom interferometer. Phys. Rev. Lett. 66 (1991) 2689--2692.

\bibitem{chapman1}
 M.S. Chapman, T.D. Hammond, A. Lenef, J. Schmiedmayer, R.A. Rubenstein,
 E. Smith, D.E. Pritchard, Photon scattering from atoms in an atom
 interferometer: Coherence lost and regained,
 Phys. Rev. Lett. 75 (1995) 3783--3787.

\bibitem{davidovic:JPhysA:2012}
 M. Davidovi\'c, A.S. Sanz, M. Bo\v{z}i\'c, D. Arsenovi\'c,
 Coherence loss and revivals in atomic interferometry: A quantum-recoil
 analysis, J. Phys. A 45 (2012) 165303(1--17).

\bibitem{asanz-bk2}
 A.S. Sanz, S. Miret-Art\'es, A Trajectory Description of Quantum Processes.
 II. Applications, Springer, Berlin, 2014.

\bibitem{talbot-jcp}
 A.S. Sanz, S. Miret-Art\'es, A causal look into the quantum
 Talbot effect, J. Chem. Phys. 126 (2007) 234106(1--11).

\bibitem{davidovic-talbot}
 M. Davidovi\'c, D. Arsenovi\'c, M. Bo\v zi\'c, A.S. Sanz,
 S. Miret-Art\'es, Should particle trajectories comply with the transverse
 momentum distribution?, Eur. Phys. J.-Spec. Top. 160 (2008) 95--104.

\end{thebibliography}

%\end{document}
%\endinput

\end{document}